\documentclass{IEEEtran}
\usepackage{cite}
\usepackage{amsmath,amssymb,amsfonts}
\usepackage{algorithmic}
\usepackage{graphicx}
\usepackage{textcomp}
\usepackage{subfig}
\usepackage{multirow}

\usepackage{bm}
\makeatletter
\AtBeginDocument{\DeclareMathVersion{bold}
\SetSymbolFont{operators}{bold}{T1}{times}{b}{n}
\SetMathAlphabet{\mathrm}{bold}{T1}{times}{b}{n}
\SetMathAlphabet{\mathit}{bold}{T1}{times}{b}{it}
\SetMathAlphabet{\mathbf}{bold}{T1}{times}{b}{n}
\SetMathAlphabet{\mathtt}{bold}{OT1}{pcr}{b}{n}
\SetSymbolFont{symbols}{bold}{OMS}{cmsy}{b}{n}
\renewcommand\boldmath{\@nomath\boldmath\mathversion{bold}}}
\makeatother

\def\BibTeX{{\rm B\kern-.05em{\sc i\kern-.025em b}\kern-.08em
    T\kern-.1667em\lower.7ex\hbox{E}\kern-.125emX}}

\begin{document}

\title{Driver-Intention Prediction with Deep Learning: Real-Time Brain-to-Vehicle Communication}
\author{{Niloufar Alavi},
{{Swati M. Shah}},
{{Rezvan Alamian}},
{and Stefan M. Götz}}

\maketitle

\begin{abstract}
Brain--computer interfaces (BCIs) allow direct communication between the brain and electronics without the need for speech or physical movement. Such interfaces can be particularly beneficial in applications requiring rapid response times, such as driving, where a vehicle's advanced driving assistance systems could benefit from immediate understanding of a driver's intentions. This study presents  a novel method for predicting a driver's intention to steer using electroencephalography (EEG) signals through deep learning. A driving simulator created a controlled environment in which participants imagined controlling a vehicle during various driving scenarios, including \textit{left} and \textit{right} turns, as well as \textit{straight} driving. A convolutional neural network (CNN) classified the detected EEG data with minimal pre-processing. Our model achieved an accuracy of 83.7\% in distinguishing between the three steering intentions and demonstrated the ability of CNNs to process raw EEG data effectively. The classification accuracy was highest for \textit{right}-turn segments, which suggests a potential spatial bias in brain activity. This study lays the foundation for more intuitive brain-to-vehicle communication systems.
\end{abstract}


\section{Introduction}
\label{sec:introduction}
Communication is an essential part of our lives. It is our way of connecting to the outer world and expressing ourselves. We may communicate with people or machines. Advances in technology and science integrate smart devices and machines into daily routines. We can use verbal, gestural, or other ways of communication to transmit our message to the machine \cite{jia2020multimodal}. Brain--computer interfaces (BCI) offer immediate communication through the brain and bypass previous modes that typically involve peripheral muscles, including speech \cite{Brain-computer2002}. Direct communication can simplify conveying a message and allow severely disabled  to do so \cite{attallah2020bci}.
BCIs use a wide range of brain signals \cite{kim2011point}. Invasive microelectrodes and epidural electrodes can detect signals even down to the individual neuron level \cite{buzsaki2004large}; noninvasive surface electrodes in electroencephalography (EEG) in turn measure electrical sum potentials of many neurons \cite{teplan2002fundamentals}. Write channels can further add bidirectional functionality \cite{sombeck2022characterizing,chen2022wireless,hughes2020bidirectional}. Although invasive methods have shown good results, e.g.,  in the control of prostheses \cite{velliste2008cortical,abiri2019comprehensive}, the risk of damage and infections associated with brain surgery, ethical aspects, and the gradual deterioration of recorded signals remain as drawbacks \cite{2012braincomputer}. Consequently, noninvasive techniques are highly preferable.

Beyond the use of EEG in medical diagnosis, BCIs have become a key application \cite{secondMeeting}. EEG signals detected concurrently with certain tasks allow the identification of correlations and patterns \cite{heim2025real,Workshop2006bci}. Such correlations and patterns can  serve as user commands; motor mu waves or specific event-related potentials may provide a physiological basis, but deep-learning approaches can even be naive to such relationships and find patterns with a sufficient quantity of training data and manage the high variability of neural signals as well as recording noise \cite{aggarwal2022has,detectionofsteering2016, kim2011point, lotte2007classificationreview,saha2020intra,goetz2014novel,li2022detection,ma2025rethinking,goetz2022isolating}.
The concurrent development of advanced machine-learning techniques can identify even small nonlinear relationships in raw high-dimensional data without reduction or aggregation \cite{tai2020impact, chen2014big, shokri2015privacy,panch2018artificial}.
Convolutional neural networks (CNN), adopted and adapted from image processing, can operate on pre-processed EEG data, which used to be the standard until most recently \cite{dose2018end,bashivan2015learning}, or directly on raw EEG signals \cite{petoku2021object,zhao2019multi, schirrmeister2017deep, tang2017hidden}.

Driving vehicles is a well-known example of a cognitively challenging task that requires many functions to operate the machine as well as managing traffic, safety, and the actual act of getting from point A to B. Accordingly, driving has stimulated massive research in simplifying it. Whereas full autonomous driving appeared to be the near future of individual transportation, its complexity has led to more attention on simplifying driving and the development of a large number of more basic or advanced driving assistants.
However, none of these systems currently takes into account the mental state or the intention of the driver when making decisions or its ability to rapidly judge situations \cite{hernandez2018eeg}. Instead, these systems exclusively rely on sensors and conventional user  \cite{chang2022driving}. The key to designing an intelligent driving assistance system, according to numerous researchers, however, is the ability to recognize driving intention \cite{haghani2021applications, xing2019driver}.

The goal of this research is to detect a driver’s or lead passenger's intention so that the machine, i.e., a car, can perform the detail work and low-level control. Previous work tried to estimate the steering intention of a driver through a combination of several sensors, such as electromyography (EMG) data, pose tracking, and EEG \cite{detectionofsteering2016, vecchiato2022eeg}. However, additional sensors are unpractical. Steering intention detection with EEG alone has concentrated on feature detection and pre-processing of the brain signal \cite{lakany2005steeringwheelchair, steeringtiming2013gheorghe}. Typical methods achieve accuracy levels of half to two thirds for three alternatives (\textit{left}, \textit{right}, and \textit{straight}) \cite{ikenishi2008classification}.
In contrast to existing work, this study bases the classification only on the EEG data and further achieves above-80\% accuracy for three categories (turning \textit{right}, turning \textit{left}, and driving \textit{straight}). Moreover, the classification is based on deep learning algorithms; no pre-processing is required, which facilitates real-time BCI control in future projects.

\section{Methods}
\subsection{Research Design}
\begin{figure}[t]
\centering
\includegraphics[width=0.9\columnwidth]{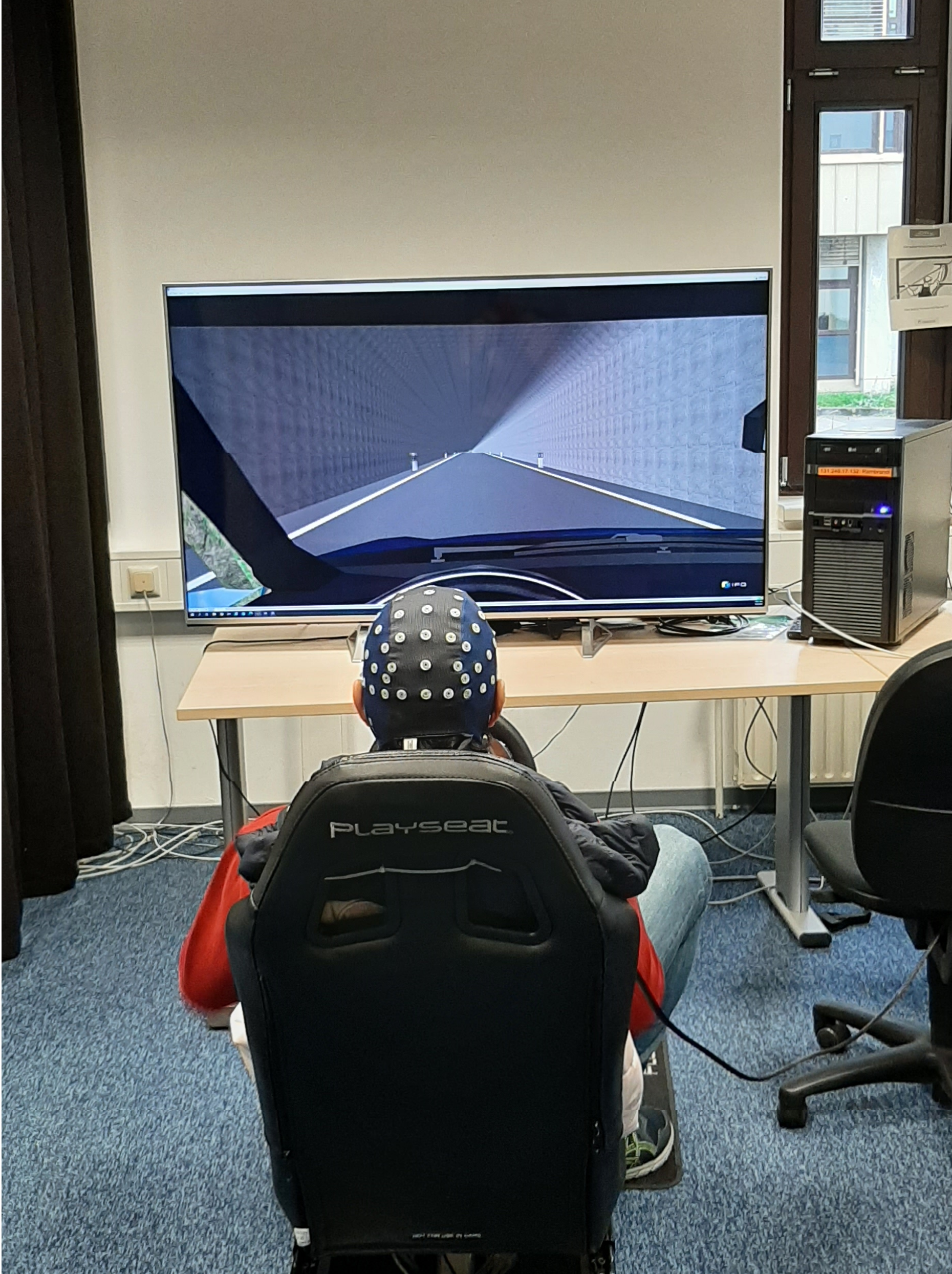}
\caption{Experimental setup of the driving simulator.
}
\label{fig_datacollection}
\end{figure}

\begin{figure} [t]
    \centering
  \subfloat[\label{cam-a}]{%
       \includegraphics[width=0.48\linewidth]{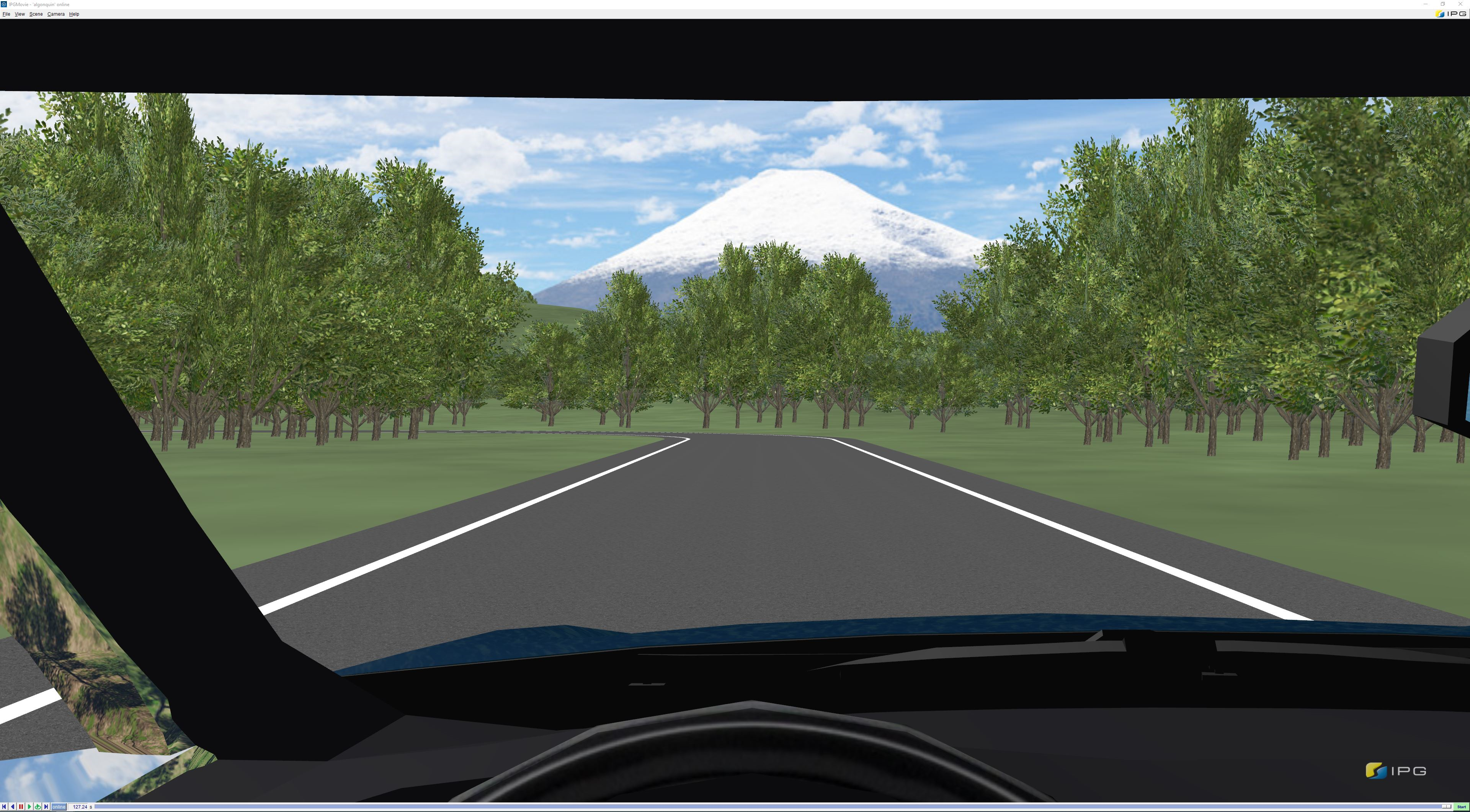}}
    \hfill
  \subfloat[\label{cam-b}]{%
        \includegraphics[width=0.48\linewidth]{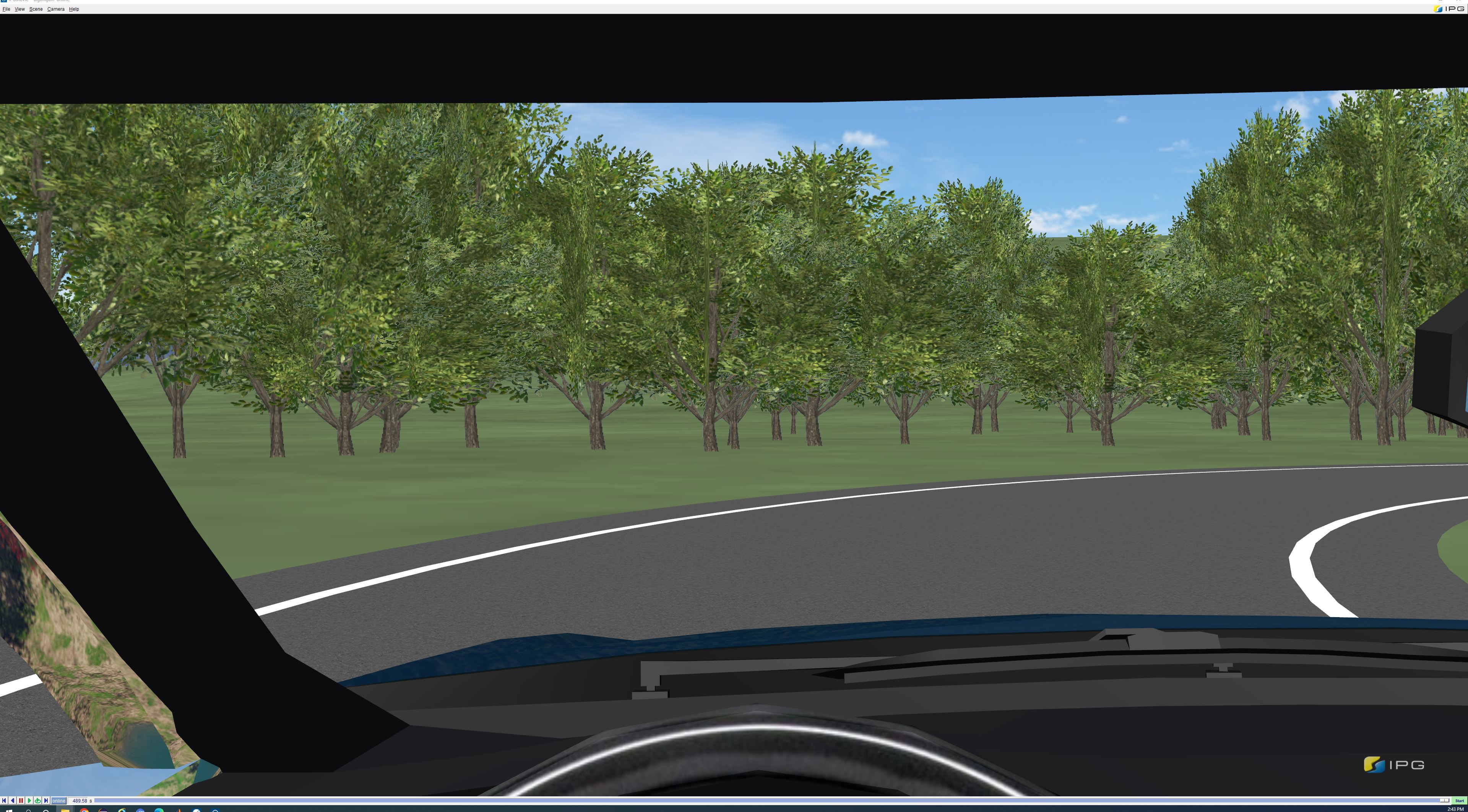}}
    \\
  \caption{Driver's view for (a) \textit{straight} driving and (b) turning. }
  \label{fig:camera} 
\end{figure}

\begin{figure}[t]
\centering
\includegraphics[width=0.99\columnwidth]{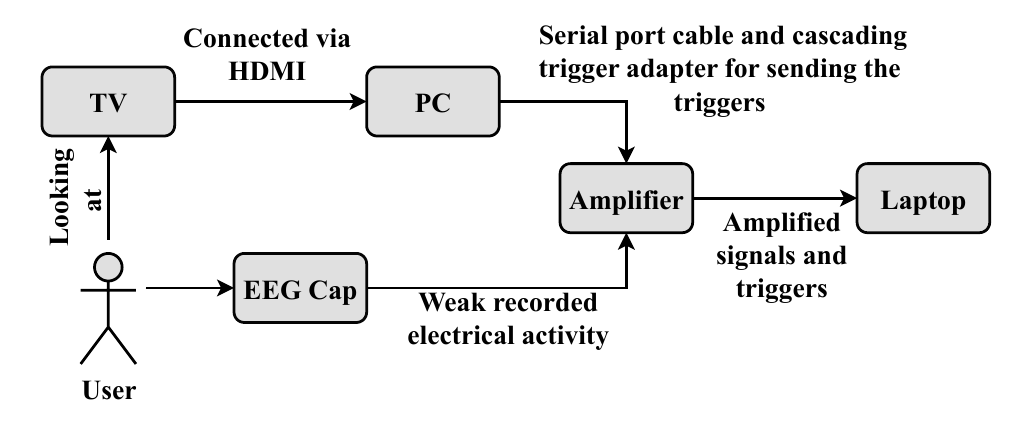}
\caption{Configuration of the signal flow in the experimental setup.}
\label{fig_Diagram}
\end{figure}

We designed a test track in IPG CarMaker (IPGMovie 9.1.1). The designed road scenario consisted of 92 segments, including \textit{left} and \textit{right} turns as well as \textit{straight} segments. The turn segments were randomly distributed; all other segment were \textit{straight}. At the beginning of each segment, a trigger was sent to the EEG software (ANT EEGo) via a serial-port cable. In the simulation, the camera position was set to an approximate driver's perspective for stronger immersion. Subjects sat 1.5 meters from a 42-inch TV screen (Fig.\ \ref{fig_datacollection}). The scenery was kept similar along the road to avoid other features affecting the EEG signal. Figure \ref{fig:camera} illustrates examples from the scenario. Road turns became visible a few seconds before participants reached them. Each complete round with a car took around 14 minutes, including three breaks (one minute each). We repeated the scenario for the participants between two and six times. The steering wheel and the pedals of the driving simulator were not used during this experiment, as the participants were supposed to only imagine making turns.
Figure \ref{fig_Diagram} illustrates the design of the experiment.

\subsection{Participants}
The participants were recruited by advertisement in social media groups and  were all students at the University of Kaiserslautern (1 male, 9 females; mean age = $28.33 \pm 6.56$, range 22--44 years). The instructions were written in English, and the participants were all English speakers with various first languages. They were right-handed with normal or corrected-to-normal vision and without any neurological or psychological disorders. Participants performed written informed consent and were compensated for their time. 
\subsection{Procedure}
The participants were asked to switch off their phones and electrical devices or put them on flight mode.
After consenting and demographics forms, subjects read the detailed instructions and could ask questions. The instructions told them to imagine that they were the driver and were controlling the car with their minds. The task was not to move or motor imagery of moving the steering wheel on the simulator, but to control the steering of the car on the screen. When they reached a turning point, they had to focus on turning in that direction and when a turn ended, they had to focus on driving \textit{straight} again. 

They were also informed about the break time, and that they should try to remain still and not get distracted during the task. Then the Waveguard gel EEG cap with 64 EEG channels 
was placed on their head according to the 10--20 system. The impedance was ensured to be below 20~k$\Omega$. The EEG data was recorded with a sampling frequency of 512~Hz.

\subsection{Analysis}
\subsubsection{Data Training}
The EEG data files were processed and analyzed in Python 3.9.1 with the Tensorflow library (version 2.8.0). No filtering or more intense pre-processing steps prepared subsequent classification. We neither removed blinking-induced eye or muscle artifacts to let the deep learning algorithms disregard  unimportant features. 
Two overlapping windows from each trigger location were extracted: Samples 700–1700 (1.36–3.32 s) and Samples 750–1750 (1.46–3.42 s), which increased accuracy by 3–4\%. Following segmentation, samples below the 10th percentile and above the 90th percentile were eliminated to remove outliers. The resulting data was normalized to zero mean and unit variance before processing by a 1-dimensional convolution neural network (CNN).  All scaled epochs across 64 channels were flattened into a single sample, and corresponding labels (\textit{straight} = 0, \textit{left} = 1, \textit{right} = 2) were assigned and saved in a database for model generation. During training and validation, data samples and labels were loaded, shuffled, and split into training (70\%) and validation (30\%) sets. The scenario naturally contained more \textit{straight} segments. Therefore, training samples were balanced using SMOTE \cite{chawla2002smote} to ensure equal representation of each class and then re-shuffled and fed to the neural network. For each segment, the sample tensor dimension was (1, 500, 64), resulting in a total training input tensor of (4278, 500, 64). The network contained convolution, max pooling, convolution, average pooling, flatten, dropout, and dense layers.
Hyperparameter tuning maximized accuracy and minimized loss through trial and error. Notably, we avoided the rectified linear unit (ReLU) activation function and zero-padding due to known low classification accuracy in certain contexts \cite{jiang2022adaptive, iqbal2022improving}.
We used the Adam optimizer with a learning rate of $2\times10^{-5}$ and categorical cross-entropy for loss calculation, appropriate for our integer output classes (0 for \textit{straight}, 1 for \textit{left}, and 2 for \textit{right}). The network was trained over 80 iterations with a batch size of 32. The best model was selected based on peak accuracy, and validation samples were evaluated using this model. The model was not tested on a held-out test set of unseen participants/sessions. A confusion matrix was derived from the actual and expected validation labels with four fundamental elements---true positives, true negatives, false positives, and false negatives. 
We used accuracy as the ratio of correct predictions versus all predictions, precision measurement for each class as the ratio of true positive predictions versus all positive predictions, recall as the ratio of true positive predictions vs. all instances of that class, and the F1 score as the harmonic balance between accuracy and comprehensiveness \cite{guerrero2021principal, powers2020evaluation}. 

\subsubsection{Visualization}
We additionally processed the EEG data for the purpose of visualization using the MNE library in Python 3.10.
To focus on the relevant brain activity, we band-pass-filtered the EEG data to 0.1–40 Hz for noise reduction and re-referencing with common-average referencing was performed to reduce the influence of common noise sources. We eliminated  artifacts, such as eye blinks, muscle activity, and electrical interference, in an artifact-rejection step with independent component analysis (ICA) to separate mixed sources of brain activity and artifacts into their independent components. 
Finally, we corrected the baseline such that the pre-stimulus baseline period ($-200$~ms until the trigger) had a mean of zero to eliminate any remaining gradual shifts in the EEG signals over time, i.e., from segment to segment. The segments were trimmed to 3~s after the trigger.
These processed data form the basis for  topographic-map plotting for each class.
We furthermore evaluated the Welch's estimate of the power spectral density as an average of periodograms over successive blocks to compensate the nonstationary nature of EEG \cite{von2010finding, welch1967use}.

\section{Results}
One participant did not finish the session and their data were excluded. The classification was carried out for all remaining participants altogether as a group model, and then on an individual basis for three random participants to identify the influence of subject variability. Table \ref{table:confmat} provides the multi-class confusion matrix based on the validation dataset for the general model as well as one of the individual models. The general model achieved an accuracy of 83.7\%.

\begin{figure}[h] 
    \centering
  \subfloat[\label{1a}]{%
       \includegraphics[width=0.49\linewidth]{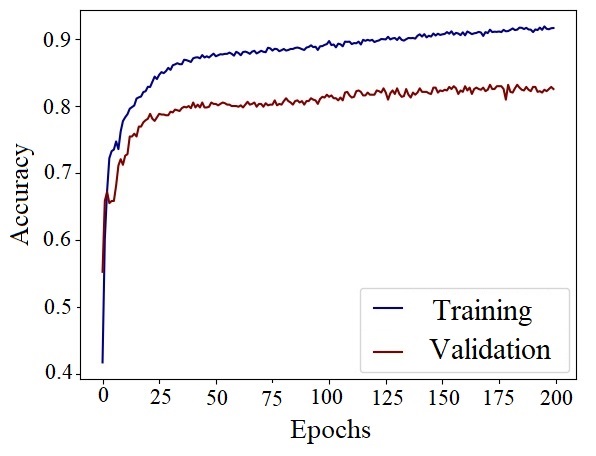}}
    \hfill
  \subfloat[\label{1b}]{%
        \includegraphics[width=0.49\linewidth]{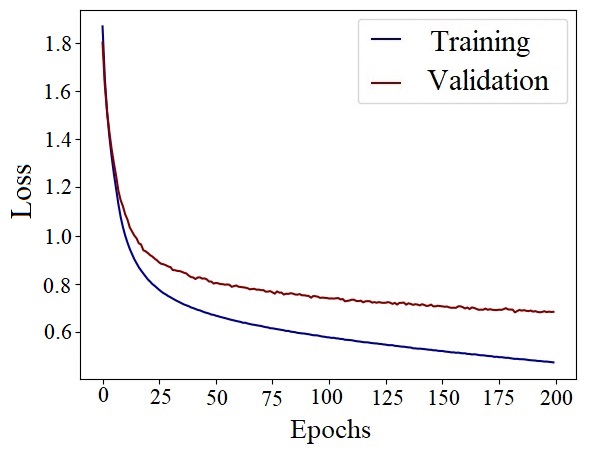}}
    \\
  \caption{Accuracy and loss for the generic model.}
  \label{fig:acc and loss general} 
\end{figure}

\begin{table}[!t]
\caption{\textbf{Confusion matrices for the generic model and individual models}}
\label{table:confmat}
\setlength{\tabcolsep}{3pt}

\begin{tabular}{|p{50pt}|p{45pt}|p{40pt}|p{40pt}|p{40pt}|}
\hline
\multicolumn{2}{|c|}{} & \multicolumn{3}{c|}{\textbf{Predicted Values}} \\
\cline{3-5}
\multicolumn{2}{|c|}{} & \textbf{Straight} & \textbf{Left} & \textbf{Right} \\
\hline

\multirow{3}{*}{\textbf{All subjects}} 
& Straight & 395 & 56 & 24 \\
& Left     & 30  & 187 & 2 \\
& Right    & 35  & 7   & 209 \\
\hline

\multirow{3}{*}{\textbf{Subject 1}} 
& Straight & 52 & 13 & 8 \\
& Left     & 6  & 23 & 0 \\
& Right    & 5  & 2  & 29 \\
\hline

\multirow{3}{*}{\textbf{Subject 2}} 
& Straight & 64 & 9  & 13 \\
& Left     & 2  & 30 & 1 \\
& Right    & 7  & 1  & 39 \\
\hline

\multirow{3}{*}{\textbf{Subject 3}} 
& Straight & 71 & 13 & 7 \\
& Left     & 3  & 21 & 3 \\
& Right    & 5  & 1  & 41 \\
\hline

\end{tabular}
\end{table}

Table \ref{table:confmat} indicates which scenarios were misclassified more than other ones. The \textit{straight} segments were classified correctly in 83\% of the cases. The misclassified data were distributed unevenly between the other two classes with a tendency to the \textit{left} class. The falsely classified \textit{left} and \textit{right} segments were mostly detected as \textit{straight}. The performance metrics in Table \ref{table:para} indicate that the  \textit{right} class has the highest scores, followed by \textit{straight}.

\begin{table}[!t]
\caption{\textbf{Performance Metrics for Each Class}}
\label{table:para}
\setlength{\tabcolsep}{4pt}

\begin{tabular}{|p{45pt}|p{40pt}|p{40pt}|p{40pt}|p{40pt}|}
\hline
\multicolumn{2}{|c|}{} & \textbf{Precision} & \textbf{Recall} & \textbf{F1 Score} \\
\hline

\multirow{3}{*}{\textbf{Categories}}
& Straight & 0.858 & 0.832 & 0.845 \\
& Left     & 0.748 & 0.854 & 0.797 \\
& Right    & 0.889 & 0.833 & 0.860 \\
\hline

\end{tabular}
\end{table}

\begin{figure} [h]
    \centering
  \subfloat[\label{topo-a}]{%
       \includegraphics[width=0.33\linewidth]{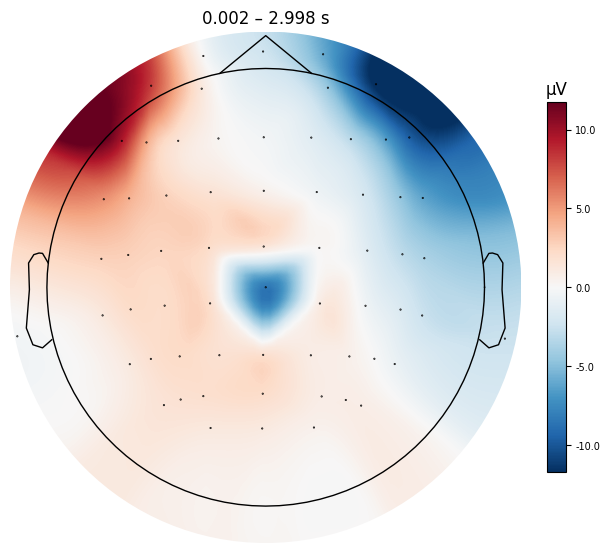}}
    \hfill
  \subfloat[\label{topo-b}]{%
        \includegraphics[width=0.33\linewidth]{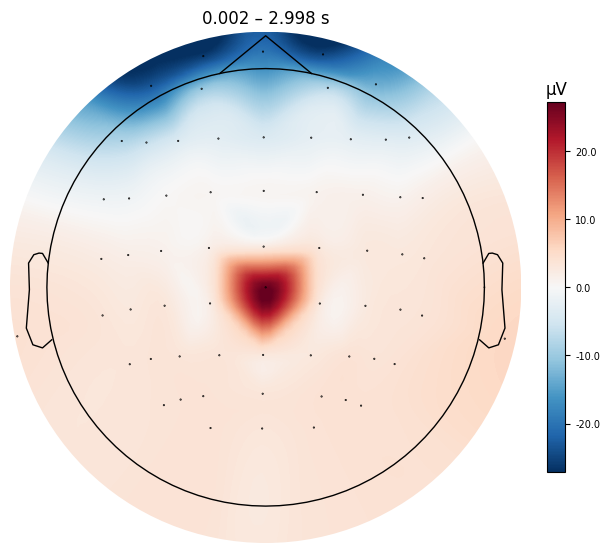}}
    \hfill
  \subfloat[\label{topo-c}]{%
        \includegraphics[width=0.33\linewidth]{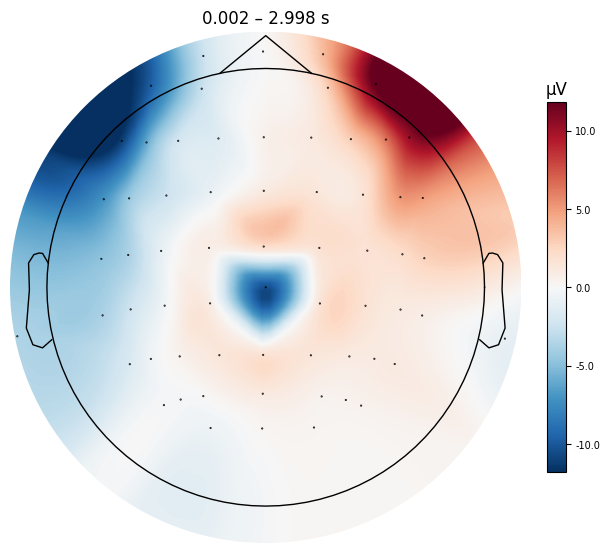}}
    \\
  \caption{ Topographic plots of average brain activity within the first 3 s after the trigger occurrence for (a) \textit{left}, (b) \textit{straight}, and (c) \textit{right}.}
  \label{fig:topographicplot} 
\end{figure}

\begin{figure} 
    \centering
  \subfloat[\label{evoke-a}]{%
       \includegraphics[width=0.49\linewidth]{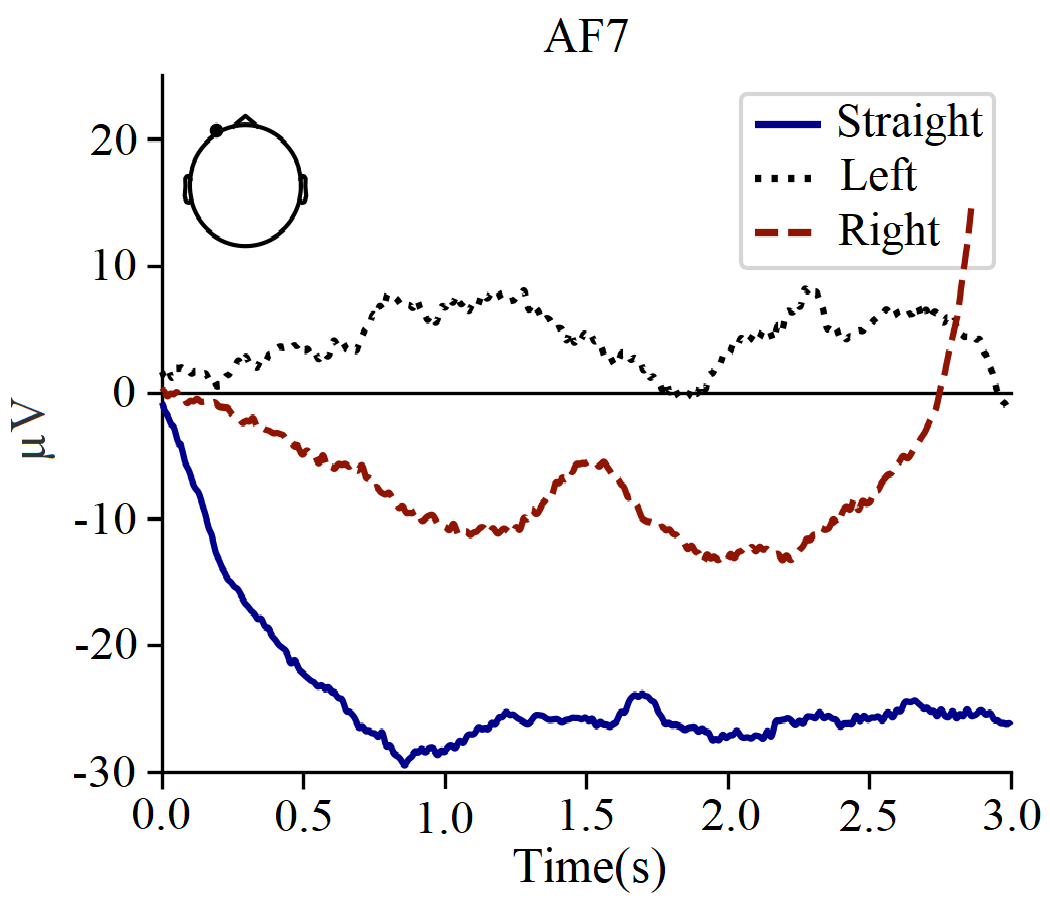}}
    \hfill
  \subfloat[\label{evoke-b}]{%
        \includegraphics[width=0.49\linewidth]{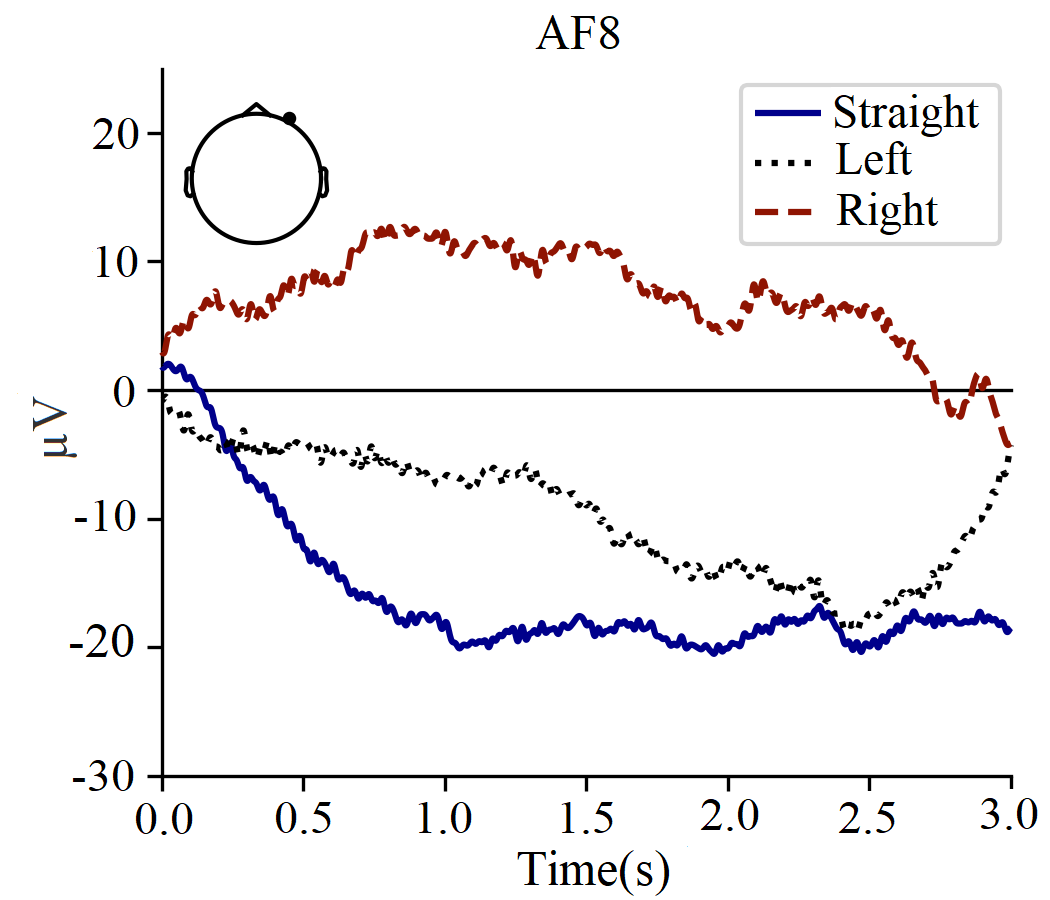}}
    \\
  \caption{Comparison of evoked responses: Plots illustrate the distinct evoked responses across different conditions for channels (a) AF7, and (b) AF8.}
  \label{fig:averageplot} 
\end{figure}

The topographic maps distinguish the three classes (Fig. \ \ref{fig:topographicplot}). During steering to the \textit{left} and \textit{right}, the frontal lobe in the corresponding hemisphere is activated more than other areas. The obvious lateral component strongly depends on the direction. Parts of frontal and parietal lobes around the central sulcus showed the highest level of activation in the \textit{straight} condition. Figure \ref{fig:averageplot} graphs the evoked potentials in Channels AF7 and AF8. The evoked potential responses vary visibly between \textit{left} and \textit{right} conditions. The polarity changes correspond to the side of the selected electrode. The pattern of activity in the \textit{straight} condition does not change with polarity but is slightly lower in the AF7 electrode.

\begin{figure}[h]
  \centering
  \includegraphics[width=0.9\columnwidth]{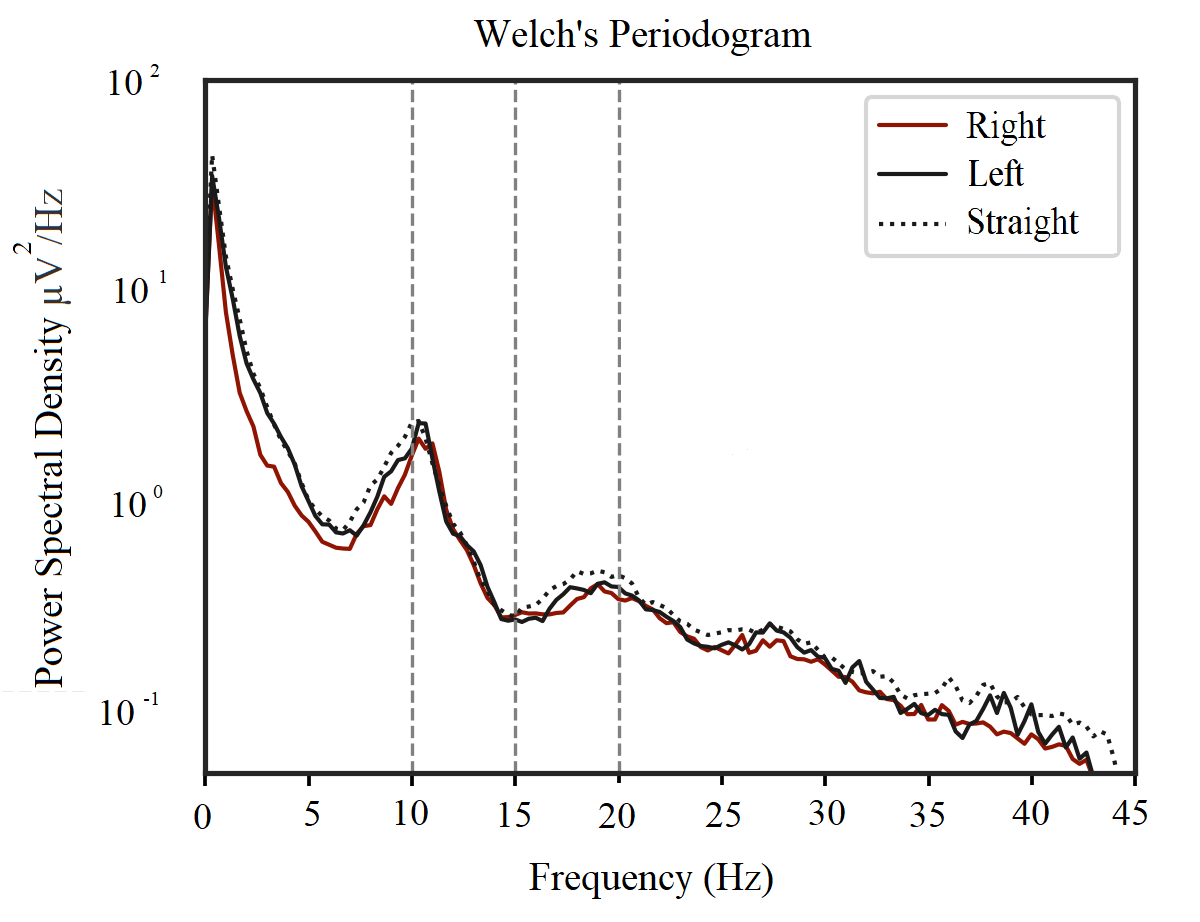}
  \caption{Power spectral density using Welch's method calculated based on the data derived from the C4 electrode.}
  \label{fig:welch}
\end{figure}

Figure \ref{fig:welch} illustrates the power spectra for the three conditions. The alpha band (8--13 Hz) consistently demonstrates a prominent peak at approximately 10 Hz. The power in the beta band (13--30 Hz) varies between conditions. For the frequencies beyond the beta range, the power spectrum gradually decreases.

\section{Discussion}
\subsection{Model Size and Variability}
The aim of this study was to predict the driver's intention of steering as one of the three categories: \textit{right}, \textit{left}, and \textit{straight} using raw data and feeding them to a convolutional neural network. We trained the model for the combined data gathered in experiments and separately for three individual data sets. We had sufficient participants and segments to use one flexible overall model and achieve high accuracy.  In return, we used a larger model than previous studies with individual models and more iterations for training. We expected a loss of accuracy compared to individual modeling as the network cannot exploit individuality anymore but could exceed previous work \cite{li2024comparison, zhang2023subject}. 

An important factor that should be considered is the variability of EEG data over time. Brigham and Kumar demonstrated that the accuracy of identifying subjects using EEG data decreases as the time between recording sessions increases and suggests that it may be useful to either refresh the training database with new data after subject identification or to use some adaptive classification approaches \cite{brigham2010subject}. Nevertheless, Maiorana et al. \cite{maiorana2015permanence} demonstrated a better rate of identification  data from distinct sessions in the training phase, as we did in our experiment. This strategy would minimize the specificity influence of the sessions that are used for the training phase and would confirm the persistence of EEG features \cite{maiorana2015permanence}.

\subsection{Sensitive Intervals and Spectral Characteristics}
It is worth mentioning that in order to reach this accuracy, epochs were extracted from 1.36 to 3.32 s after the trigger points. Epochs from zero to two seconds after the trigger points did not produce a good outcome. This low performance of the immediate time after the trigger is expected as related studies have also used epochs in a similar range such as six-second epochs (including moments before and after the steering) \cite{ikenishi2008classification}, 2.5 s before and 2.5 s after the events \cite{vecchiato2022eeg}, and four seconds after the start of steering \cite{steeringtiming2013gheorghe}.

The Welch’s periodogram of the EEG data shows a high peak around 0.5 Hz in all conditions. This peak could be related to physiological activity or artifacts such as electrode drift or breathing-related movements. With the frequency increase from delta (0 Hz –- 4 Hz) to theta
(4 Hz –- 8 Hz), we observe a decrease in power, which is typical in EEG recordings and
consistent with $1/f$ spectral behavior of brain signals \cite{demanuele2007distinguishing}. 
The peak in alpha (8 Hz -- 13 Hz) range can be seen for all the different conditions and suggests
similar relaxation levels during each task. Differences between
conditions are more distinct in the beta band (13 Hz -– 30 Hz), where the power spectra for imagining \textit{left} and going \textit{straight} demonstrate slightly higher power compared to imagining turning \textit{right}. The spectral difference between the conditions suggests differential involvement of motor or cognitive processes, with more cortical activation during the \textit{left} and \textit{straight} imagery tasks. In the gamma (30 Hz – 45 Hz) range, the power spectrum gradually decreases and there is no significant peak in this range. The power spectra for the three conditions follow a similar trend across all frequency bands, but their most differences are within the beta band.

Additionally, the patterns seen in the topographic maps may not merely reflect brain activity related to imagining turning. They likely additionally capture highly correlated eye movements. The strong signals over the frontal areas are typical of activity generated when the eyes move \cite{ronca2024optimizing}. Eye activity is hard to separate but on the other hand also a fair contribution to the signal and a means to the intended end of high-level vehicle steering. When participants watched the car turning to the \textit{right}, the rest of the turn appeared in their \textit{right} visual field and they looked at the \textit{right} side more often. The same applied to the \textit{left} turns. This gaze behavior is in line with studies showing that drivers tend to look toward the future path of the vehicle when steering \cite{land1994we, tuhkanen2021visual}.

\subsection{Spatial Bias}
Brain data for turning \textit{right} were classified more accurately than the others (Tables \ref{table:confmat} and \ref{table:para}). The \textit{right}--\textit{left} bias was unexpected, as the number of samples for \textit{straight} segments was almost double that of the other segments. We expected a higher accuracy for \textit{straight} segments and rather an equal accuracy for \textit{left} and \textit{right} turns. One possible explanation could be the fact that spatial attention is not evenly divided into the two hemispheres, and the bias caused by attention leans more towards the \textit{right} \cite{takio2014influence}.
Other research also supports this finding, but the results are limited to right-handed people, as the subjects were right-handed \cite{takio2013visual}.
Hence, the directional accuracy bias is unlikely a result of our setup.
The formation of spatial biases can be influenced by handedness-related factors \cite{bareham2015does}.
Thus, handedness is supposedly the reason behind the higher accuracy in the \textit{right} segments, and the results may vary for left-handed people, where the accuracy might be higher for the \textit{left} turns. However, more evidence is required for such a conclusion and further research can be carried out in the future.

Another notable point is that the \textit{straight} segments that were not correctly classified were mistaken for \textit{left} segments in most cases. However, the turning segments were most often misclassified as \textit{straight}. Accordingly, the network can distinguish \textit{left} and \textit{right} turns quite well, but the distinction between a turn and a \textit{straight} segment is not equally strong. A possible reason is that although the instructions clearly said that the participants had to focus on turning and going \textit{straight}, some of them only concentrated on turning, meaning that during the \textit{straight} segments, they struggled with imagining driving \textit{straight}.

In conclusion, the aim of this research, which was to predict the driver’s steering intention was successfully fulfilled. Deep learning is a promising field that has the potential to significantly improve the classification accuracy of raw data.

\bibliographystyle{IEEEtran}

\bibliography{Refs}
\end{document}